# Using Applied Behavior Analysis in Software to help Tutor Individuals with Autism Spectrum Disorder

Antonio Ugando
Georgia Institute of Technology
Atlanta, USA
tugando@gatech.edu

**ABSTRACT**
There are currently many tutoring software systems which have been designed for neurotypical children. These systems cover academic topics such as reading and math, and are made available through various technological mediums. The majority of these systems were not designed for use by children with special needs, in particular those who are diagnosed with Autism Spectrum Disorder. Since the 1970's, studies have been conducted on the use of Applied Behavior Analysis to help autistic children learn [1]. This teaching methodology is proven to be very effective, with many patients having their diagnosis of autism dropped after a few years of treatment. With the advent of ubiquitous technologies such as mobile devices, it has become apparent that these devices could also be used to help tutor autistic children on academic subjects such as reading and math. Though the delivery of tutoring material must be made using Applied Behavior Analysis techniques, given that ABA therapy is currently the only form of treatment for Autism Spectrum Disorder endorsed by the US Surgeon General [2], which further makes the case for incorporating it into an academics tutoring system tailored for autistic children. In this paper, we present a mobile software system which can be utilized to tutor children who are diagnosed with Autism Spectrum Disorder in the subjects of reading and math. The software makes use of Applied Behavior Analysis techniques such as a Token Economy system, visual and audible reinforcers, and generalization. Furthermore, we explore how combining Applied Behavior Analysis and technology, could help extend the reach of tutoring systems to these children.

**ACM Classification Keywords**
J.4 [**Computer Applications**]: Social and Behavioral Sciences; K.4.2 [**Social Issues**]: Assistive technologies for persons with disabilities.

**INTRODUCTION**
Autism is a developmental disorder which currently affects 1 in 59 children in the US [3]. The typical traits of autism include repetitive behaviors, trouble initiating eye contact, sensory issues, and overall challenges with social interactions and communication. The onset of this disorder in some cases begins to appear at around 18 months of age, upon which a regression in developmental skills are witnessed (e.g. the child loses language which was previously acquired, and/or the child becomes socially detached). In these cases, the child was meeting developmental milestones prior to the onset of the disease. For this reason, there appears to be a consensus among researchers, which consider Autism Spectrum Disorder to be a combination of both genetic predisposition and epigenetic environmental factors [4]. Some children who suffer from Autism Spectrum Disorder are also labeled as intellectually disabled. This is usually found in severe cases of autism, although most autistic children are very smart, though their social barriers make them appear to suffer from Intellectual Disability, which upon closer analysis is not the case. The social interaction challenges which affect autistic children place them at a disadvantage when compared to their neurotypical counterparts. This particular deficit later leads to other deficits in skills such as speech and communication [5], which could further affect academic progress later on in life. For decades, the common understanding was that autistic children were unteachable, primarily due to these characteristics. It was not until a study was published in 1987 by Ivar Lovaas, at the University of California, Los Angeles [8], which showed that you could indeed teach autistic children, though the technique used to teach them has to be very structured, in particular utilizing what is now referred to as Applied Behavior Analysis. Through this study, Lovaas was able to debunk this stereotype which was associated with autistic children, ushering in a new era of learning for these children. For over four decades now, Applied Behavior Analysis has been used as a mechanism in treating children who suffer from Autism Spectrum Disorder. This form of therapy has been thoroughly studied, and has been shown to be a very successful method in treating the disorder [6]. The applications of ABA therapy are broad, and can be used to help these children advance in: social, developmental, and academic skills, to name just a few. This form of therapy is currently performed by clinical psychologists, in particular trained Board Certified Behavior Analysts (BCBAs), who sit through intense daily sessions with an autistic child for many hours, usually around 25 or more per week [6], in order to help the child meet certain predefined milestones. These milestones can range from social to developmental to academic in nature. Through this treement, the child is able to progressively overcome these social, developmental, and academic deficits. The primary concept behind ABA is to reinforce good behavior and actions by using a rewarding system, and refrain from promoting bad behavior and unwanted actions by not issuing any rewards. An example of this would entail a therapist asking a child to identify a truck from a list of images, and if the child correctly chooses the truck, the therapist responds with congratulatory excitement, thus reinforcing the correct action taken by the child. Another type of reward may include allowing the child to watch their favorite cartoon for one minute, or play with a toy which they enjoy for the same amount of time. Other rewarding systems include a Token Economy System [7], in which a child is provided with what looks like a miniature BINGO board, and gets to fill the board with tokens which are earned while successfully progressing throughout a session. Once the board is filled with tokens, the child gets to exchange the tokens for a reinforcer: cartoon, toy, playtime, etcetera. These rewarding techniques are used thoroughly throughout ABA therapy, with the optimal goal of reinforcing desired behaviors in the child while slowly eliminating unwanted ones. Given the success which has been associated with ABA therapy as identified through research, it is considered by clinicians as the primary tool which is available for treating Autism Spectrum Disorder. This is why a strive was made to introduce ABA techniques into the system which is being presented in this paper, so that tutoring sessions in reading and math could be delivered in an effective manner. The intent of this software is to fill a gap which exists within tutoring systems tailored for individuals with Autism Spectrum Disorder. This is accomplished by providing the child with mechanisms which can be used to reinforce good and desired behaviors, such as through the use of rewarding systems. Such rewarding systems include the Token Economy System [7], as well as visual and audible praising. The child is notified of their progress through the use of this Token Economy System, which provides them with a measurable means of knowing when to expect a reward, which in the case of this system is an open source cartoon. The system also ensures that the student is generalizing concepts, by presenting them within varying contexts (e.g. various animals eating, similar summation using different objects), while keeping track of any progress which is made. The system runs on either an Android phone or tablet, and thus is optimized for touch. A Board Certified Behavioral Analyst was consulted during the design of this system, in order to ensure that the final product resulted in a viable and effective tool for tutoring autistic children.

**The Case for ABA**

In the past 20 years, 7 long-term, large-scale controlled studies have demonstrated that children who receive more than 25 hours per week of Applied Behavior Analysis for more than 1 year make tremendous gains, with some participants achieving functioning within the average range for their age [6]. Out of these long-term studies, the one which is most referenced by many Board Certified Behavior Analysts happens to also be the first of its type. This first study, which evaluates the long-term effects of Applied Behavior Analysis on children with Autism Spectrum Disorder was conducted in 1987 by Ivar Lovaas, at the University of California, Los Angeles [8]. In this study, Lovaas was able to show through empirical evidence that ABA is an effective form of therapy for treating Autism Spectrum Disorder. The study included an experimental group of clinically labeled autistic children who were administered ABA therapy for 40 hours or more per week, and a control group with the same label which received only 10 hours or less per week of this form of therapy. The therapy sessions were administered in a one-to-one fashion, meaning that one therapist was assigned to one child throughout each session. The therapy lasted for 2 or more years, throughout which data was collected pre and post treatment. The results were staggering, after 2 years of therapy, 47% of children in the experimental group fell within the average level IQ of their neurotypical peers, while only 2% of the children in the control group had similar outcomes. To make matters even more interesting, McEachin, Smith, and Lovaas, published a follow-up study in 1993, in which they reevaluated these same children from the study which was published in 1987. What they found offers much hope to caregivers of children suffering from this terrible disease, 8 out of 9 children continued to progress, and did not show any signs of regression [9]. This study took the child mental health industry by storm, because it proved for the first time, that if an autistic child is administered intensive ABA therapy for at least 2 years, at a very early age, there chances of eventually losing the label of autism are high. Though having an autistic child reach this milestone is not an easy task, primarily because ABA therapy is quite expensive, and requires many years of intervention. On average, the cost of issuing ABA therapy to a child with autism ranges between $40,000 and $60,000 per year [3]. The good news is that many states in the United States of America, have passed legislation mandating that insurance providers include ABA therapy coverage in their clients' policies [10]. For those who do not have private medical insurance, ABA therapy is covered through the US government sponsored Medicaid program [11]. Seeing how the rate of autism cases have drastically increased over the last few decades, currently standing at 1 in 59 children affected by the disease [3], it is reassuring to see the US government taking a stance in the fight against this debilitating disease. There is currently hope for families with children on the autism spectrum, and thanks to grassroots efforts such as Autism Speaks, autism awareness is more prevalent today than it was just a few years ago.

**RELATED WORK**
Having observed Board Certified Behavior Analysts perform ABA therapy, they mostly rely on sketches and pictures throughout a session. The only time they rely on technology is when they are issuing their patient a reinforcer, such as a cartoon to watch via a tablet, or a game to play via a similar medium. The actual learning tasks and milestones which are associated with a patient are all issued through the use of hard copy material. An attempt to replace this manual process was introduced in 2017 by Kyle Miller, at Georgia Institute of Technology, Atlanta [12]. Miller's system replaces the process of using hard copy material with a web based application, in which you are able to upload material tailored to a specific patient. Miller's system is a huge step in the right direction when it comes to adding Educational Technology to the treatment of Autism Spectrum Disorder, though it is not designed for tutoring per se. His system only accepts static media, such as images, and thus it does not adapt to the patient's learning trends. For instance, his system does not help a patient generalize a concept, which is a skill that is critically missing from autistic children [13]. An example of such a system would present the patient with a couple of examples, say of a cat and a cow eating (with the answer given), and then present the patient with a third example of a dog eating, upon which the patient is asked "What is the dog doing?", and if the patient answers incorrectly, then the application provides the answer, followed by another example of an animal eating, as well as the question, until the patient is able to generalize the concept. Another ABA technique which is missing from Miller's system is the use of a Token Economy System [7], which is considered a strong reinforcer of positive responses from the patient, as it provides the patient with a visible queue of the remaining steps which are required to obtain a reward: toy, video game, cartoon, etcetera. A similar application to that which was introduced by Miller was developed by Alexis Bosseler and Dominic W. Massaro at University of California [14]. This application uses an animated face, which prompts an autistic child with questions related to vocabulary words, upon which the child has to select an answer. The application is called Baldi, and the author states that the facial representation was introduced in order to simulate a more human like interaction between the application and the patient. This application uses rewards in the form of a happy, sad, and confused emoji face, in order to provide the patient with feedback on the outcome of their selection. Though this rewarding mechanism which the application uses does not conform to the types of reinforcement seen in an ABA session, as was illustrated via examples earlier. Baldi can also be considered a predated system in today's world, given that it is confined to a PC with a mouse, with no support for touchscreen enabled devices. Overall, failure to adopt learning techniques such as ABA makes it close to impossible to teach autistic children any form of academic material. This is because children who are diagnosed with Autism Spectrum Disorder learn in a different manner than neurotypical children [15].

**THE SOLUTION**
There is currently a void of tutoring systems which utilize Applied Behavior Analysis techniques to help teach autistic children academic subjects such as reading and math. The lack of such systems causes caregivers and Board Certified Behavior Analysts to allocate money and time to tutoring autistic children in these academic subjects. The time and money which is spent by caregivers on these tutoring sessions could be automated with technology, given that it is delivered using the scientifically backed ABA approach. Similarly, time spent by a BCBA in tutoring a child with Autism Spectrum Disorder could be redirected to helping the child learn how to speak, or become potty-trained, etcetera. With the advent of mobile devices, such as smartphones and tablet computers, the time is ripe for taking advantage of these technologies, in order to help deliver tutoring material to autistic children via ubiquitous devices such as these, while integrating a research-backed approach such as ABA. The commoditization of these mobile devices lends itself to opportunities such as these, which could help caregivers and BCBAs redirect their resources to helping a child meet other milestones. These devices also offer a much more convenient means of delivering a tutoring system than say for instance the Baldi system which was introduced in 2003 by Alexis Bosseler and Dominic W. Massaro, at University of California [14]. The system which they introduced was pre mobile smart devices, and thus was stationary and required a personal computer along with a mouse. Refreshingly, we are no longer constrained to these limitations in technology. Therefore, the system which is being presented leverages mobile device technology, in particular the Google Android mobile platform. This system helps fill a gap which currently exists within tutoring systems tailored for individuals with Autism Spectrum Disorder. The system relies on the use of techniques which are found in the research backed Applied Behavioral Analysis therapies, in order to successfully issue tutoring sessions in reading and math. These techniques include providing the child with mechanisms which can be used to reinforce good and desired behaviors, such as through the use of rewarding systems. Such rewarding systems include a Token Economy System, as well as visual and audible praising.

The child is notified of their progress through the use of this Token Economy System, which provides the child with a measurable means of knowing when to expect a reward (e.g. cartoon, toy, etcetera). The system also ensures that a child is generalizing concepts, by presenting them within varying contexts (e.g. various animals eating, similar summation using different objects), while keeping track of any progress which is made.

**Design**

In Figure 1, we see a depiction of the Home screen to the application. A user is able to access this screen by selecting the Home item in the navigation menu which is found at the bottom of the screen. This navigation menu is conveniently located in the same location throughout the application, which helps facilitate quick and easy access to the main screens that make up the application. Consideration was taken into making the design intuitive and minimalistic, in order to ensure that it delivers a user-friendly experience.

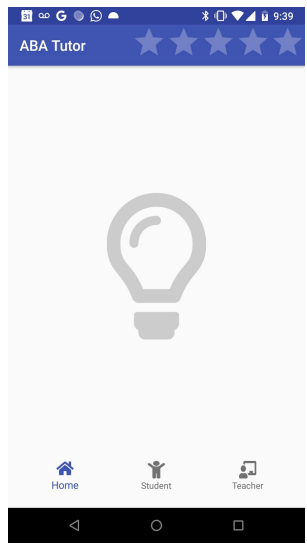

**Figure 1. Home screen**

In Figure 2, we can see a depiction of the Teacher screen. This screen can be utilized by either a Board Certified Behavior Analyst or a student's caregiver, in order to enter tutoring material into the application. When trying to access this screen, the user is presented with a math challenge question, which helps restrict access to the screen only to teachers. This screen shows a list of example tutoring questions which have been entered into the application, along with the associated image and answer for each. The teacher is able to enter tutoring data by selecting the circular button containing the embedded plus sign, which hovers at the bottom-right of the screen. Upon selection of this button, the teacher is presented with the screen in Figure 3, which allows them to enter tutoring data. Once the data is entered, it will appear in the list which is found on the screen in Figure 2. A teacher could later edit this information by selecting the item from within the list, upon which they will be taken to the screen in Figure 3.

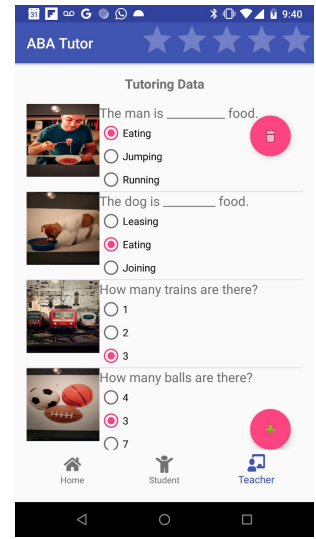

**Figure 2. Teacher screen**

Figure 3 represents the screen which is used to enter tutoring data into the application. In order to ensure that the student is generalizing information which is being presented to them, the teacher is forced to enter at least two questions for each classification, when entering data via this screen. For instance, if the teacher wants to present material related to the verb "eating", he/she will need to enter at least two examples in which this verb is used. Similarly, if the teacher would like to present the math concept of counting, they would need to enter at least two examples which depict the notion of counting. Generalization is a concept which autistic children have trouble mastering, and thus it is a common skill which is targeted throughout Applied Behavior Analysis [13].

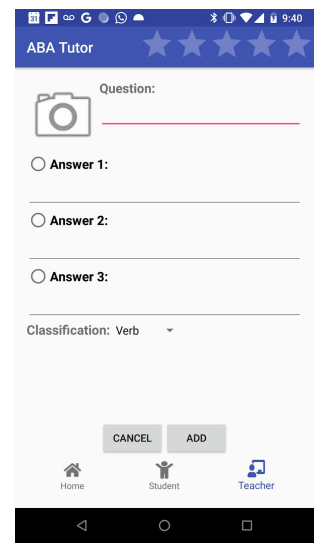

**Figure 3. Teacher add media screen**

In Figure 4, we are presented with the Student screen, which is used to deliver all tutoring material that is being presented to the student. Upon selecting the Student item in the navigation menu which is found at the bottom of the screen, the tutoring workflow process is initiated, after which the application starts presenting tutoring material to the student. A sample representation of the data which will be presented to the student can be seen in this figure. The image which is seen in the figure represents the image on which the question and answer is based. If a student selects the correct answer to the question which is being asked, the application presents the student with a dialog congratulating them as well as playing an uplifting sound effect (see Figure 5). Whereas if their selection is not correct, the application presents the student with a dialog which shows the correct answer, and it plays a somber sound effect. The application immediately presents the student with another question similar to the one they answered incorrectly, in order to ensure that the student is able to generalize the concept. In addition, the application reinforces the correct answer which was chosen by the student by issuing a token to the student (represented as stars at the upper-right), and letting them know how many tokens he/she has gained so far. Once the student reaches 5 tokens, they are exchanged for a cartoon which is played for them from within the application (see Figure 6). This method of rewarding is called a Token Economy system [7], and is considered a very strong reinforcer, because students quickly learn that the better and faster they complete the tutoring session, the faster they can trade their tokens in for a desired reward.

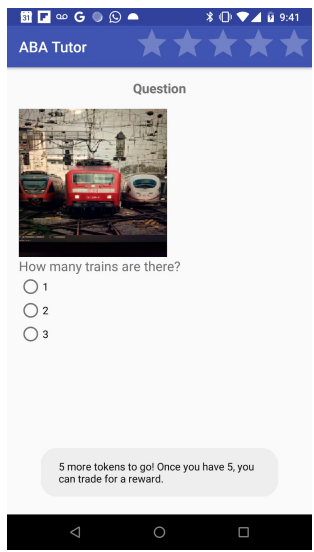

Figure 4. Student screen

In Figure 5, we can see a representation of a positive reinforcer. Positive reinforcers are used throughout Applied Behavior Analysis in order to acknowledge to a child that an action or behavior which they have taken is correct. In order to be of any use, these positive reinforcers must be delivered immediately, so that the child learns to associate the correct action with the positive acknowledgement.

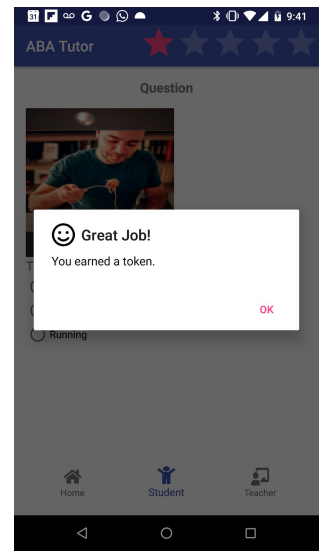

Figure 5. Student visual reinforcer

Figure 6 depicts a representation of what the internal reward would be after a student earns 5 tokens, upon which they are traded in for this particular reward. In this case, the student would be allowed to watch a cartoon which is provided to them from within the application, but only for a max length of 1 minute and 15 seconds. The student is also notified that they have earned 5 tokens, in order to help them understand how a Token Economy works. Once the student receives this reward, the tokens are reset to 0, upon which the student must answer more tutoring questions in order to receive the reward again. The cartoon which is served as a reward within the ABA Tutor application is an open source animation named Big Buck Bunny.

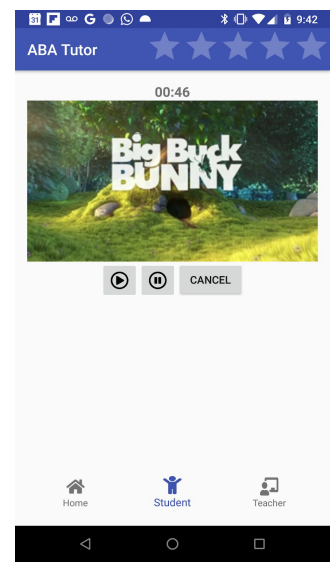

Figure 6. Reward screen

**Technology**

The ABA Tutor application was developed using Google's Android Studio IDE, in particular utilizing the Java programming language. All information is stored locally on the device, and thus it works fully offline. It should be noted that ABA Tutor operates in a fully anonymous fashion. The application is currently deployed on the Google Play store, and can be installed on any device running Android version 7.0 (Nougat) or above by searching "ABA Tutor" on the app store.

**METHODOLOGY**

A five day trial was established in order to measure user acceptability and performance. The metrics which were used during this trial include: Time of Engagement (hrs), Generalization Rate (%), and Accuracy Rate (%). Time of Engagement is a metric used to represent the amount of time in hours, in which the user engaged with the application before switching to another task. The Generalization Rate is used to measure the rate at which the subject was able to generalize a concept. For instance, if the child answered a question incorrectly, but then answered the follow-up generalization promoting question correctly, the Generalization Rate would be recorded as 1 out of 1, or 100%. The third metric recorded is Accuracy Rate, which is used to measure the rate at which the user was able to provide 5 correct answers before receiving their reward. Therefore, if the user answered 5 out of 10 questions correctly prior to receiving their reward, the Accuracy Rate would be 50%. This metric is important, as it can indirectly represent how motivated the user is in receiving the reward, and it also provides a direct representation of how well the child is learning the material.

This trial was ran on one child who is diagnosed with Autism Spectrum Disorder. Due to limitations with time, it was impossible to obtain the necessary consent which would have allowed the inclusion of more individuals in the trial.

**THE RESULTS**

Results of the trial are recorded in Figure 7. A noticeable metric in this figure is the Time of Engagement, which happens to trend upwards. In the case of this study, it was visibly noticeable that the child happened to find the reward engaging, and soon learned the mechanism to obtaining it. Another metric which trended up during the 5-day trial is the Accuracy Rate, which signifies that the child was able to learn the tutoring material. The final metric which was used to record user acceptability and performance throughout this trial is the Generalization Rate, which slightly trended up as well, which shows that the child did show some signs of generalization. It should be noted that these results can be taken with a grain of salt, given that the study involved just one child. An effort can be made to validate these results by running another trial with a larger pool of users.

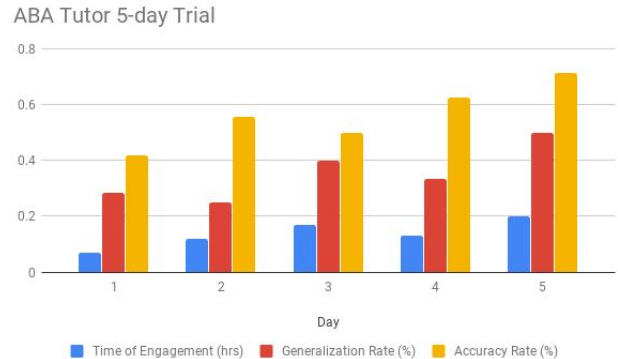

**Figure 7. ABA Tutor 5-day Trial**

**LIMITATIONS**

There are some concerns which must be considered when replacing Board Certified Behavior Analysts with a fully automated tutoring system. One of these concerns includes the assertion that Applied Behavior Analysis is being applied in an appropriate and effective manner. In the case of the ABA Tutor application which is being presented, this was mitigated by having a BCBA evaluate and approve of the system throughout its development. Another concern which poses a significant threat to the efficacy of such a system includes any social regression which may be witnessed in a child utilizing the system, given that a human is no longer delivering the tutoring session. This is a concern which was mentioned in the paper involving the Baldi system, and the authors mitigated it by introducing a human face to simulate a human tutor providing the child with instructions. Another strategy which could be considered is to limit the screen time while using the system, and thus ensuring that the child takes breaks and interacts with a human subject.

**CONCLUSION**

Although there currently does not exist any known cure for autism, Applied Behavior Analysis does offer a glimmer of hope. This form of therapy has been shown to help autistic children make extensive progress [6]. In addition, research has shown that the prognosis is bleak for children with autism who receive little to no ABA therapy [9]. By combining this form of therapy with modern technologies such as smart mobile devices, an effective tool was developed, which could help tutor autistic children on the academic subjects of reading and math. This in turn allows caregivers and Board Certified Behavior Analysts to redirect their efforts towards helping a child meet other milestones.

## FUTURE WORK

Many enhancements are considered for version 2 of the ABA Tutor application. The first consists of adding support for multiple students. The initial idea was that the student or child would provide their own personal device, though it is understood that this is not always possible, for instance in the case where a Behavior Analyst wants to run the application on their own device while serving multiple students. It is worth noting that this enhancement introduces concerns with privacy, given that user accounts would be created for primarily disabled minors on a third party device. Therefore, regulations such as FERPA and IDEA should be reviewed prior to implementing this enhancement. The second enhancement involves the inclusion of a scoring mechanism into the application, which would provide an automated way of tracking progress. A third enhancement is to make the classifications list a dynamic value, which would provide an unlimited number of generalization opportunities. Additional enhancements consist of making the buttons bigger, and thus more child friendly, and adding text to speech to the application, in order to introduce verbal prompts. A further enhancement entails providing the student with other reward options (e.g. the ability to exchange tokens for other cartoons or an external reward). An additional task which is planned is to target earlier versions of Android, which was not possible in version 1 due to a reliance on Java lambda expressions. This would allow for a wider adoption of the application, as many Android users are still running older versions of the OS. Another planned task which would further promote adoption is to target the Apple iOS platform as well. Achieving this task may entail porting the codebase to frameworks such as Xamarin or Cordova. Adding these planned enhancements to the application would broaden its reach, and thus help many more individuals along the way.

## ACKNOWLEDGMENTS

I would like to give thanks to my wife, for providing me with much needed support throughout my research and the development of this system. I dedicate this project to my daughter Shayla, who is diagnosed with Autism Spectrum Disorder, and who provides me with the drive and inspiration to work on projects such as this. May this system be a useful tool for other autistic children, their caregivers, and therapists.